\documentclass[referee,useAMS,usenatbib,usegraphicx]{mn2e}

\newcommand{\BP}{Ballesteros-Paredes}

\newcommand{\cs}{c_{\rm s}}
\newcommand{\dSFE}{$\langle\dot{\rm SFE}\rangle$}

\newcommand{\Eg}{E_{\rm grav}}
\newcommand{\Ek}{E_{\rm kin}}

\newcommand{\kms}{{\rm ~km~s}^{-1}}
\newcommand{\Lbox}{L_{\rm box}}
\newcommand{\linf}{l_{\rm inf}}

\newcommand{\Mbox}{{\cal M}_{\rm box}}
\newcommand{\Mdense}{{\cal M}_{\rm dense}}
\newcommand{\Minf}{{\cal M}_{\rm inf}}
\newcommand{\Mrms}{M_{\rm s,bgd}}
\newcommand{\Ms}{M_{\rm s}}
\newcommand{\Msinf}{M_{\rm s,inf}}
\newcommand{\Msinks}{{\cal M}_{\rm stars}}
\newcommand{\Msun}{M_\odot}
\newcommand{\pcc}{{\rm ~cm}^{-3}}

\newcommand{\Rinf}{R_{\rm inf}}
\newcommand{\sfeabs}{SFE$_{\rm abs,25}$}
\newcommand{\sferel}{SFE$_{\rm rel,5}$}
\newcommand{\sfrff}{SFR$_{\rm ff}$}
\newcommand{\tcros}{t_{\rm cros}}

\newcommand{\tsf}{t_{\rm SF}}

\newcommand{\vinf}{v_{\rm inf}}
\newcommand{\vrms}{v_{\rm rms}}
\newcommand{\VS}{V\'azquez-Semadeni}

\title[Star Formation Efficiency in
Simulations of Molecular Cloud Formation]{Dependence of the Star
Formation Efficiency on the Parameters of Molecular Cloud Formation
Simulations}

\author[Rosas-Guevara et al.] {Yetli Rosas-Guevara$^{1}$ \thanks{E-mail:
y.rosas@crya.unam.mx}, Enrique \VS$^{1}$ \thanks{E-mail:
e.vazquez@crya.unam.mx}, Gilberto C. G\'omez$^1$ \thanks{E-mail:
g.gomez@crya.unam.mx},
\newauthor and A.-Katharina Jappsen$^2$ \thanks{E-mail:jappsena@Cardiff.ac.uk}
\\
$^{1}$Centro de Radioastronom\'\i a y Astrof\'\i sica,
Universidad Nacional Aut\'onoma de M\'exico, Apdo. Postal 3-72, Morelia,
58089, M\'exico\\
$^2$School of Physics \& Astronomy, Cardiff University,
Queens Buildings, The Parade, Cardiff CF24 3AA, UK
}

\begin{document}

\maketitle

\label{firstpage}

\begin{abstract}
We investigate the response of the star formation efficiency (SFE) to
the main parameters of simulations of molecular cloud formation by the
collision of warm diffuse medium (WNM) cylindrical streams, neglecting
stellar feedback and magnetic fields. The parameters we vary are the
Mach number of the inflow velocity of the streams, $\Msinf$, the rms
Mach number, $\Mrms$ of the initial background turbulence in the WNM,
and the total mass contained in the colliding gas streams,
$\Minf$. Because the SFE is a function of time, we define two estimators
for it, the ``absolute'' SFE, measured at $t = 25$ Myr into the
simulation's evolution (\sfeabs), and the ``relative'' SFE, measured 5
Myr after the onset of star formation in each simulation (\sferel). The
latter is close to the ``star formation rate per free-fall time'' for
gas at $ n = 100 \pcc$. We find that both estimators decrease with
increasing $\Minf$, although by no more than a factor of 2 as $\Msinf$
increases from 1.25 to 3.5. Increasing levels of background turbulence
(injected at scales comparable to the streams' transverse radius)
similarly reduce the SFE, because the turbulence disrupts the coherence
of the colliding streams, fragmenting the cloud, and producing
small-scale clumps scattered through the numerical box, which have low
SFEs. Finally, the SFE is very sensitive to the mass of the inflows (at
roughly constant density and temperature), with
\sferel\ decreasing from $\sim 0.4$ to $\sim 0.04$ as the mass in the
colliding streams decreases from $\sim 2.3 \times 10^4 \Msun$ to $\sim
600 \Msun$ or, equivalently, the virial parameter $\alpha$ increases
from $\sim 0.15$ to $\sim 1.5$. This trend is in partial agreement with
the prediction by \citet{KM05}, since the latter lies within the same
range as the observed efficiencies, but with a significantly shallower
slope. We conclude that the observed variability of the SFE is a highly
sensitive function of the parameters of the cloud formation process, and
may be the cause of significant scatter in observational determinations.

\end{abstract}
 

\section{Introduction} \label{sec:Intro}

The control of the star formation efficiency (SFE) by turbulence is a central
issue in our present understanding of star formation, and currently a
topic of intense study \citep[see, e.g., the reviews by][]{MK04, MO07}. In
recent years, several groups have studied the SFE of molecular clouds
(MCs) using numerical simulations of isothermal turbulence, in which the
entire numerical box represents the interior of a molecular cloud
\citep[see, e.g., the reviews by][]{MK04, BP_etal07, VS07}. One confusing
issue is that simulations of {\it driven} turbulence seem to indicate
that the SFE {\it decreases} as the turbulent rms Mach number $\Ms$
increases \citep[e.g.,][]{KHM00, VBK03, VKB05}, while simulations of
{\it decaying} turbulence suggest that the SFE {\it increases} with
increasing $\Ms$ \citep{NL05}. This has prompted the question of how
does turbulence actually originate and behave in real MCs. To answer
this question, it has become necessary to investigate the entire
evolutionary process of MCs.

The formation of MCs by collisions of warm neutral medium (WNM) streams
has been intensely studied in recent years. A vast body of numerical
simulations has shown that moderate, transonic compressions in the WNM
can nonlinearly trigger a phase transition to the cold neutral medium
(CNM) \citep[e.g., ][]{HP99, KI00, KI02, WF00}, and that the
dense gas produced by this mechanism is overpressured with respect to
the mean WNM thermal pressure \citep{VS_etal06} and turbulent, due to
the combined action of Kelvin-Helmholz, thermal \citep{Field65} and
nonlinear thin-shell \citep{Vishniac94} instabilities
\citep{Heitsch_etal05, Heitsch_etal06}. The turbulence produced by this
mechanism is continually driven for as long as the compression lasts.

The physical scenario of MC evolution was
outlined by \citet{HBB01}, who estimated the column densities necessary
for the cloud to become self-gravitating, molecular, and magnetically
supercritical, finding them to be comparable. The one-dimensional
physical conditions in the dense gas were calculated analytically by
\citet{HP99} and \citet{VS_etal06}. A recent review of the subject has
been presented by \citet{HMV08}.

More recently, simulations including self-gravity and ``sink
particles'', which represent gravitationally collapsed objects (stars or
stellar clusters), and using finite-duration compressions in the WNM,
although lacking stellar feedback and magnetic fields, have been used to
study the evolution of the turbulent motions and of the SFE in a
self-consistent manner \citep[][hereafter Paper I]{VS_etal07}.
Concerning the velocity dispersion in the clouds, these authors found
that the turbulence is intermediate between driven and decaying, since
what decays is the driving rate of the turbulence as the inflows weaken
with time. However, they also found that the random motions are gradually
replaced by global infall motions, as the cloud begins to contract
gravitationally. Concerning the SFE, Paper I measured the masses of
dense gas $\Mdense$ and of the collapsed stellar objects $\Msinks$ in
the simulations, allowing a measurement of the SFE, defined as
\begin{equation}
\hbox{SFE} = \frac{\Msinks}{\Mdense + \Msinks}.
\label{eq:SFE_def}
\end{equation}
The resulting SFE was however too high, reaching $\sim 50$\% roughly 6
Myr after the time at which star formation (SF) had begun (denoted
$\tsf$), although this excessive SFE can possibly be attributed to the
neglect of stellar feedback in that simulation. Indeed, Paper I estimated,
using a prescription by \citet{FST94} and a standard IMF
\citep{Kroupa01}, that by 3 Myr after $\tsf$, enough massive stars would
have formed as to be able to destroy the cloud by ionization. At that
point, the SFE was $\sim 15$\%, closer to the typical values $\la 5$\%
reported observationally for full MC complexes \citep{Myers_etal86}.

One obvious possible reason for the relatively high values of the SFE
obtained in this type of simulations is the neglect of stellar feedback
and magnetic fields. However, it is also of interest to perform a more
systematic investigation of the degree of variability that the SFE could
exhibit within the scenario of Paper I, simply by varying the parameters
of the WNM compressions triggering the formation of the cloud
complex. In this paper we undertake a first approach to such task, by
varying three parameters of the WNM stream collisions modeled in the
simulations. First, we consider the inflow speed $\vinf$ and the
velocity dispersion of the background turbulence initially present in
the medium, both measured by their respective Mach numbers, $\Msinf$ and
$\Mrms$, with respect to the unperturbed WNM. Subsequently, we consider
the mass in the colliding streams $\Minf$, as determined by their radius
$\Rinf$ and length $\linf$. Since the parameter space covered by these
three parameters is already quite large, in this work we do not consider
variations in the collision angle of the streams, instead having them
collide head-on in all cases.

The paper is organized as follows. In \S \ref{sec:the_model} we describe
the numerical model and experiments. In \S \ref{sec:results} we present
our results, and in \S \ref{sec:conclusions} we present a summary and
our conclusions.

\section{Numerical model and experiments} \label{sec:the_model}


For the numerical simulations, we use the same numerical setup as that
used in Paper I, except that we now use the smoothed particle
hydrodynamics (SPH) + $N$-body code Gadget-2 \citep{Springel05} (in
Paper I we used the previous version of the code, Gadget), modified to
include random turbulence driving and sink particles according to the
prescription of \citet{Jappsen_etal05}, and including parameterized
heating and cooling, as applied in Paper I, using the fit of
\citet{KI02} to a variety of atomic and molecular cooling
processes. This cooling function causes the gas to be thermally
unstable, under the isobaric mode, in the density range $ 1 \la n
\la 10 \pcc$. We assume that the gas is all atomic\footnote{Our
``molecular clouds'' are thus only so in the sense of density and
temperature, but not of chemical composition.}, with a mean atomic
weight of 1.27. The numerical box is periodic, with size $\Lbox$. In all
cases, we use $118^3 = 1.64 \times 10^6$ particles, and set the mean
number of particles within a smoothing volume to 40. According to the
criterion of \citet{BB97}, the mass resolution is twice the number of
particles within a smoothing volume, or $\sim 80$ times the mass per
particle. The critical density for sink formation is set at $3.2
\times 10^7 \pcc$, and the outer sink accretion radius is set at 0.04 pc.

An initial turbulent velocity field of one-dimensional velocity
dispersion, characterized by its rms Mach number $\Mrms$ and applied at
scales between 1/4 and 1/8 of the box size, is added to the inflow
velocity field, in order to trigger the instabilities that render the
cloud turbulent. Note that this added turbulent velocity field is
applied by turning on the random driver for the first few timesteps of
the simulation's evolution, and that what we actually control is the
energy injection rate parameter. Thus, simulations intended to have the
same turbulence strength do so only approximately, as the flow's
response is slightly different in every realization.

The initial conditions consist of a uniform medium at $n = 1 \pcc$ and
$T = 5000$ K, in which two cylinders of length $\linf$ along the $x$
direction and radius $\Rinf$ are set to collide head-on at the $x =
\Lbox/2$ plane of the simulation (refer to Fig.\ 1 of Paper I). Note
that the cylindrical inflows are entirely contained within the numerical
box, since the boundaries are periodic. The length $\linf$ is measured
from the central collision plane, and is always shorter than the
half-length of the box, implying that a small region between the edge of
the inflows and the box boundaries is not given any velocity. This
region is partially evacuated during the subsequent evolution of the
simulations, as the gas within it tends to fill the void left by the
inflows.

Table \ref{tab:run_params} shows the various runs we performed for the
present study, indicating the relevant parameters for each one. In
addition to the inflow length, radius, and velocity, defined above,
Table \ref{tab:run_params} gives the Mach number $\Msinf$ corresponding
to $\vinf$ at the initial temperature of the gas, for which the
adiabatic sound speed is $7.54 \kms$, the one-dimensional rms Mach
number of the initial turbulent motions, $\Mrms$, the total mass
contained in the simulation box, $\Mbox$, the mass contained in the two
inflows, $\Minf$, and the mass per SPH particle, ${\cal M}_{\rm
part}$. The runs are labeled mnemonically, with their names giving, in
that orded, the inflow Mach number $\Msinf$, the background Mach number
$\Mrms$, and the inflow mass, $\Minf$.

\begin{figure}
\includegraphics[width=1.\hsize]{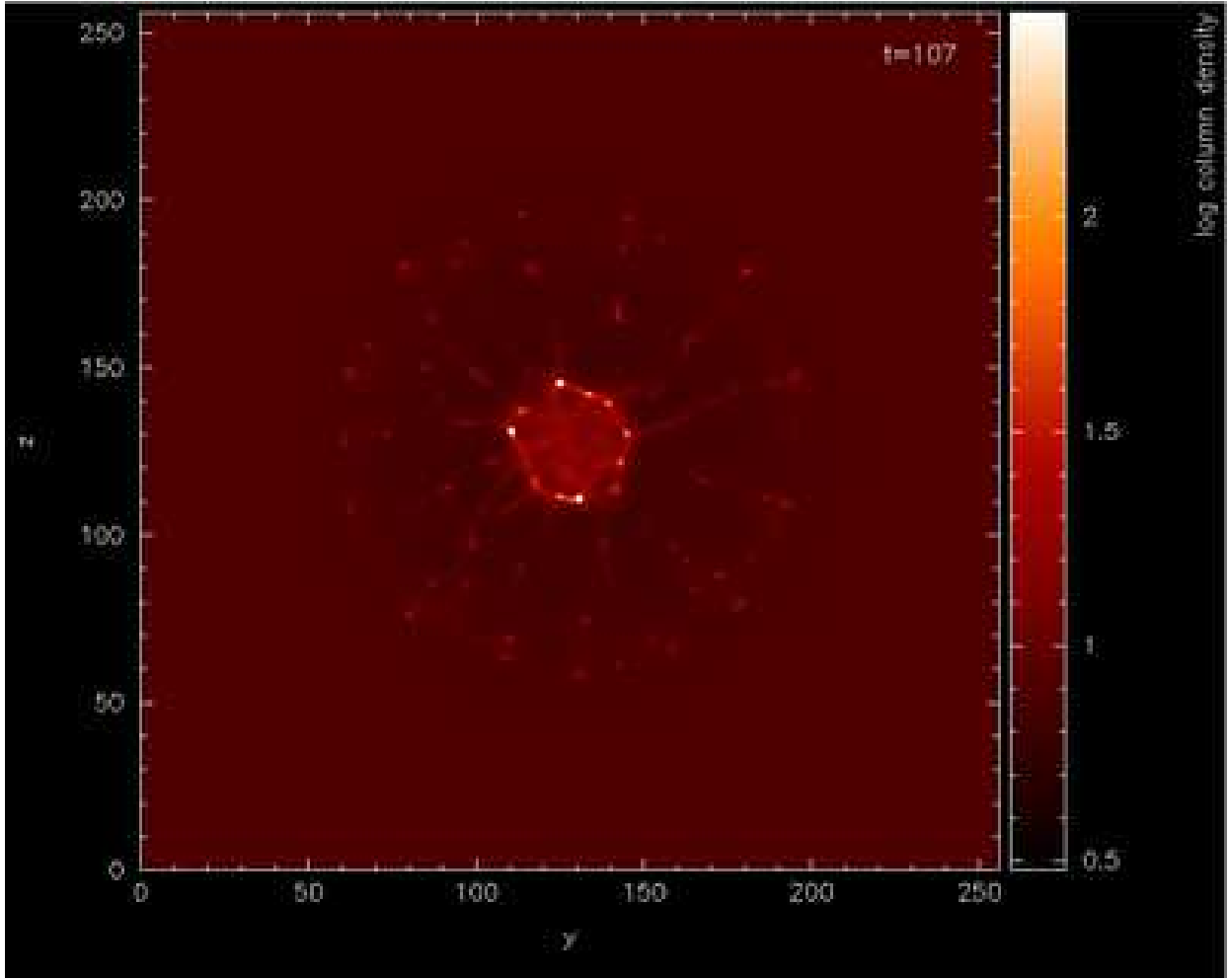}
\caption{Face-on image in projection of run Mi1-Mb.02-Ma2e4 at the time
it is beginning to form stars. Note the central ring and the radial
filaments, charactristic of this type of runs (Paper I).}
\label{fig:run4_img}
\end{figure}

Note that run Mi1-Mb.02-Ma2e4 has the same parameters as run
L256$\Delta v$0.17 from Paper I, and we take these as the ``fiducial''
set of parameters. Note, however, that these two runs are not identical
because they were performed with different codes. Gadget-2 differs in
many ways from Gadget, and in particular it takes longer timesteps,
implying that the initial forcing (used to trigger the instabilities in
the dense layer) is applied at different time intervals, and with
different random seeds. Thus, the two runs are similar only in a
statistical sense, but are not identical. For reference, in
Fig. \ref{fig:run4_img} we show a face-on view of Mi1-Mb.02-Ma2e4 at the
time when it is beginning to form stars, showing that the general morphology
it develops is similar to that of run L256$\Delta v$0.17 from Paper I
(compare to Fig.\ 4 of that paper, noting that the linear scales shown are
different in each figure).

We report the SFE as defined by eq.\ (\ref{eq:SFE_def}), with the
``dense'' gas defined as that with number density $n \ge 100
\pcc$. However, because the SFE is actually a function of time, and sink
formation begins at different times in different runs, in order to
report a {\it number} for the SFE, we estimate it in two different
ways. One is to measure the ``absolute'' SFE 25 Myr after the start of
the simulation, which we denote as \sfeabs. The other is to measure the
``relative'' SFE, which we define as the SFE 5 Myr after the onset of
star formation in the simulation, and which we denote as \sferel. This
is actually close to the ``star formation rate per free-fall time'',
SFR$_{\rm ff}$ (after SF has begun), as defined by \citet{KM05}, since
the free-fall time for gas at $n \sim 100 \pcc$ is $\sim 4.6$ Myr.

\section{Results} \label{sec:results}

\subsection{SFE vs.\ inflow velocity} \label{sec:SFE_vs_vinf}

\begin{figure}
\includegraphics[width=0.45\hsize]{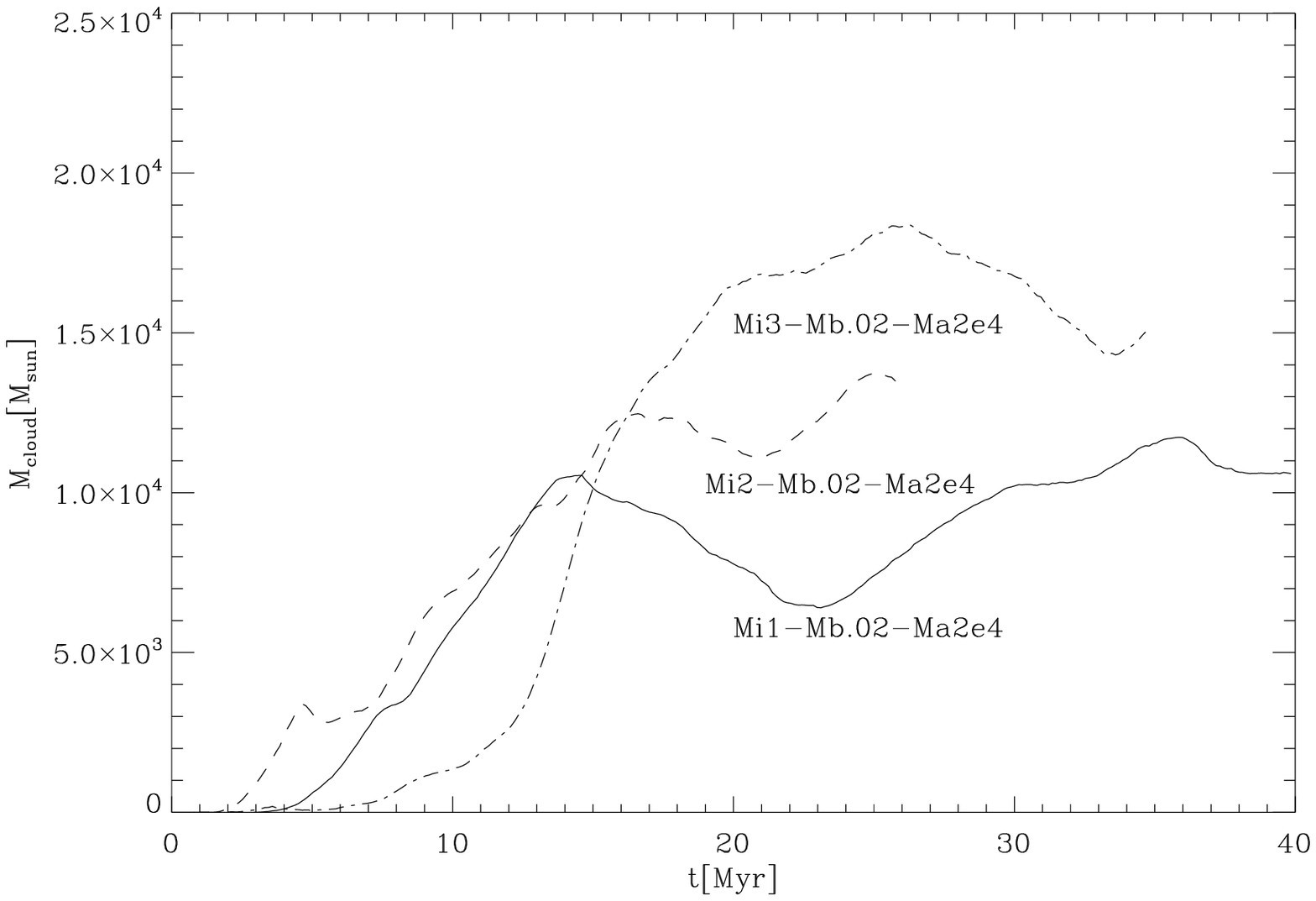}
\includegraphics[width=0.45\hsize]{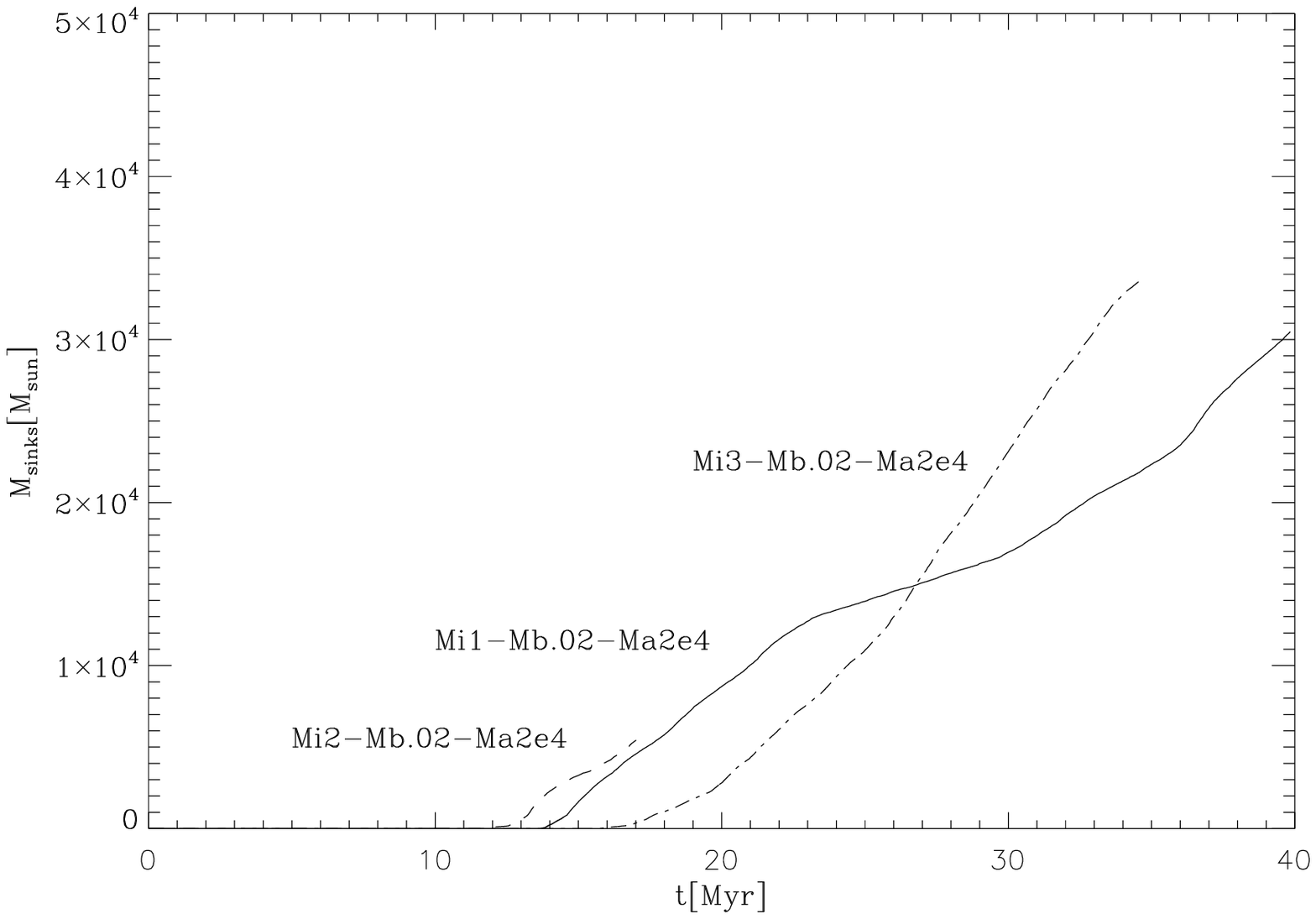}
\caption{Evolution of the dense gas mass ({\it left panel}) and sink mass
({\it right panel}) for runs Mi1-Mb.02-Ma2e4, Mi2-Mb.02-Ma2e4, and
Mi3-Mb.02-Ma2e4, which differ only by the Mach number of the inflows
(indicated by the ``Mi\#'' entry in the run's name), having
$\Msinf = 1.25$, 2.5 and 3.5 respectively.}
\label{fig:Mg_Ms_of_t_vs_vinf}
\end{figure}

\begin{figure}
\includegraphics[width=0.5\hsize]{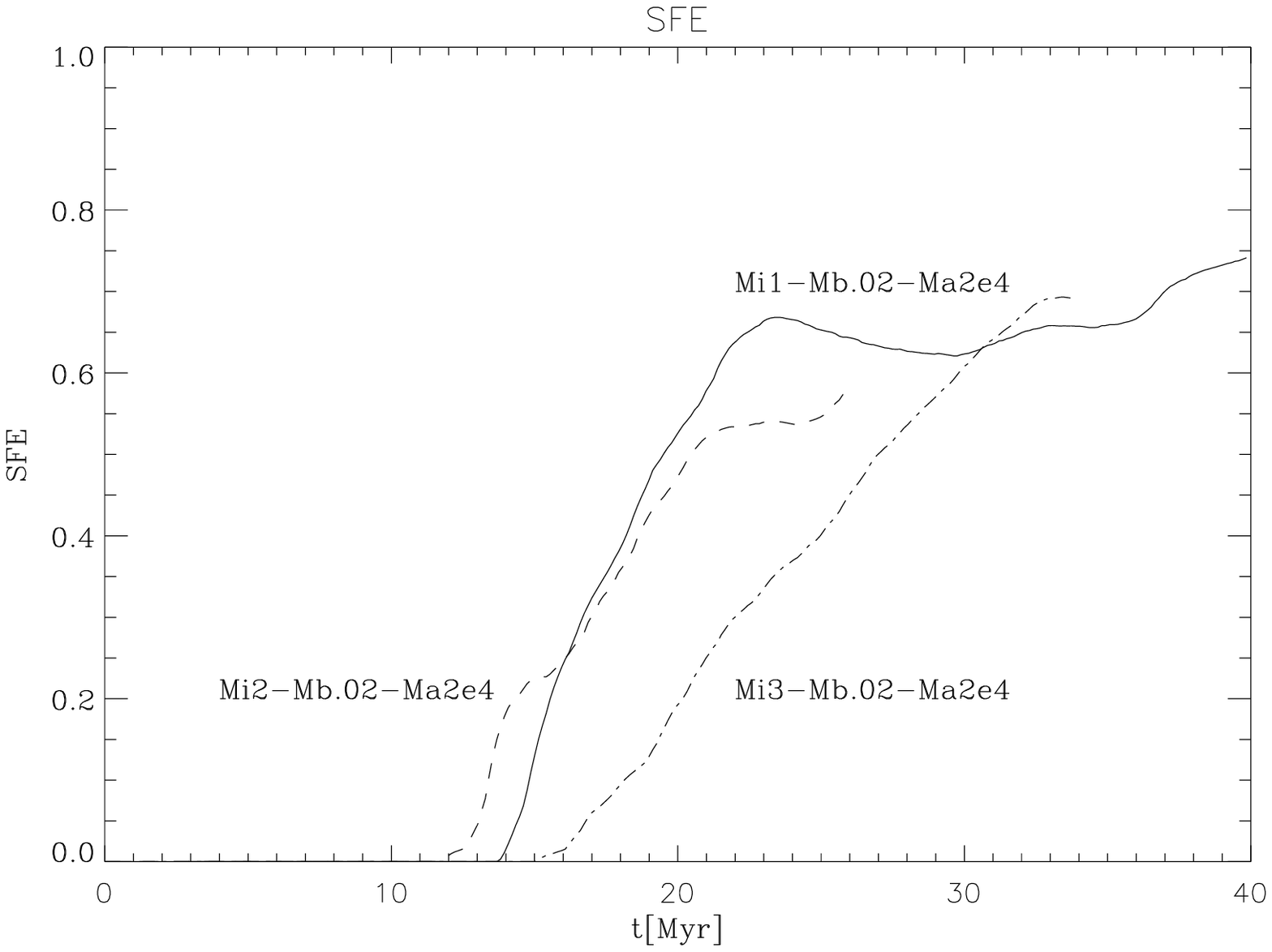}
\includegraphics[width=0.5\hsize]{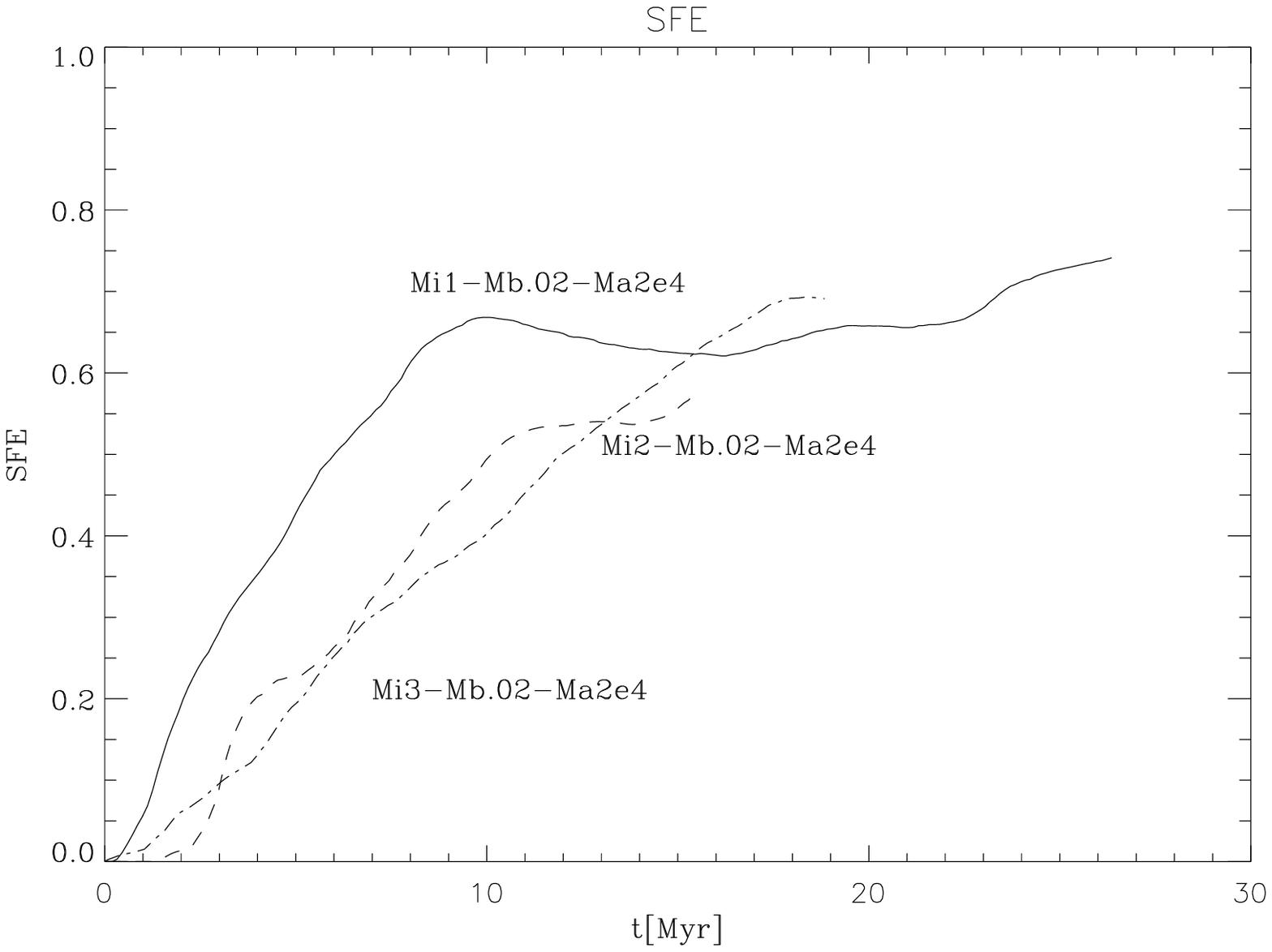}
\includegraphics[width=0.5\hsize]{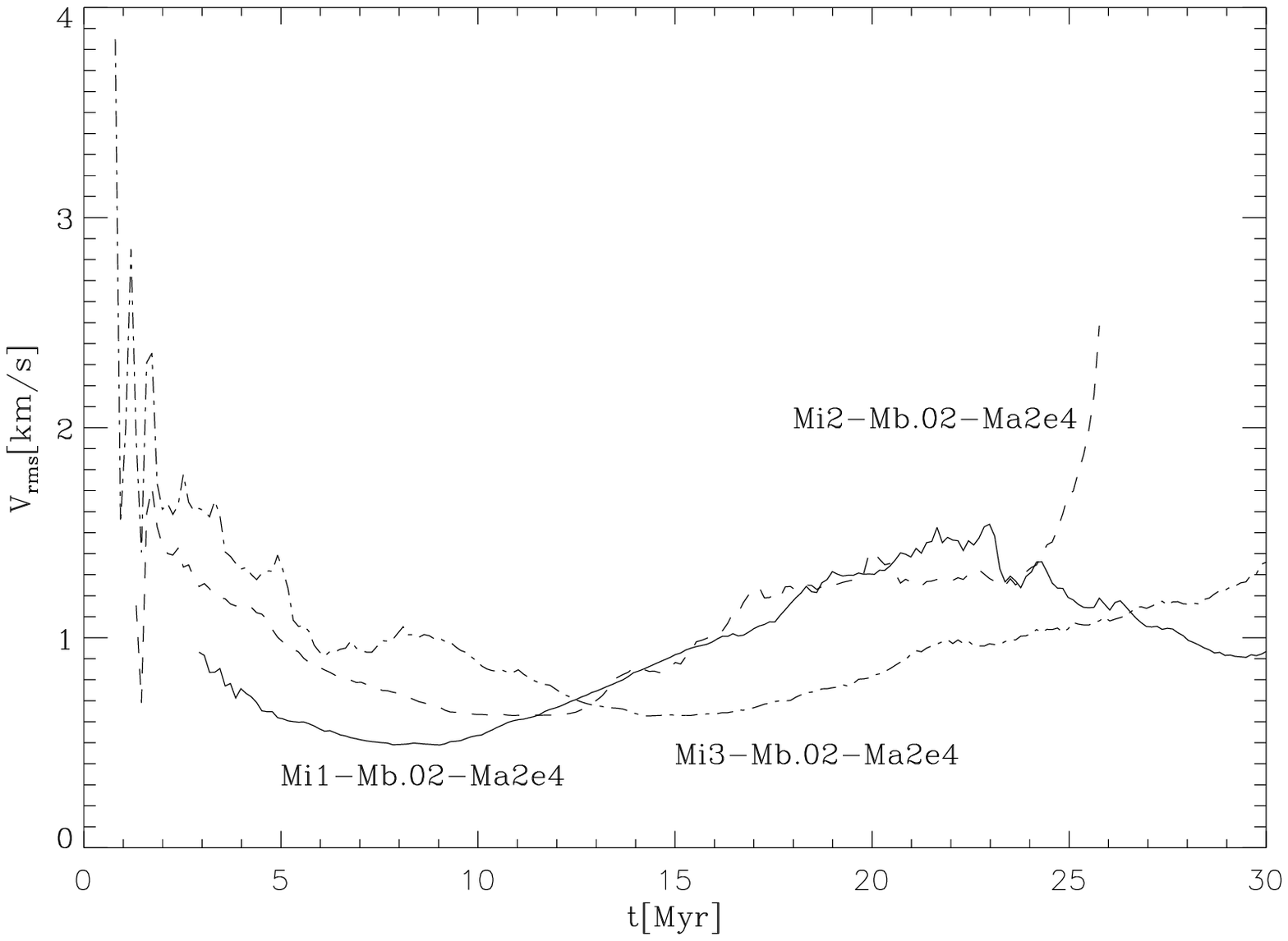}
\caption{{\it Top panels:} Evolution of the ``absolute'' SFE (\sfeabs,
{\it left}), shown out to 25 Myr after the start of the runs, and the
``relative'' SFE ({\it right}), shown from the onset of sink
formation, for runs Mi1-Mb.02-Ma2e4, Mi2-Mb.02-Ma2e4, and
Mi3-Mb.02-Ma2e4. \sferel\ is the value of this curve at
a relative time of 5 Myr. {\it Bottom panel:} Evolution of the
one-dimensional velocity dispersion perpendicular to the direction of
the inflows for the same three runs.}
\label{fig:SFE_of_t_vs_vinf}
\end{figure}

In this section, we consider the dependence of the SFE on the inflow
velocity of the colliding streams, $\vinf$. Figure
\ref{fig:Mg_Ms_of_t_vs_vinf} shows the evolution of the dense gas mass
({\it top panel}) and the sink mass ({\it bottom panel}) for runs
Mi1-Mb.02-Ma2e4, Mi2-Mb.02-Ma2e4, and Mi3-Mb.02-Ma2e4. These runs have
all parameters equal, except for the speed of the inflows (cf.\ Table
\ref{tab:run_params}), which are varied from $\Msinf = 1.25$ in
Mi1-Mb.02-Ma2e4 to $\Msinf = 3.5$ in Mi3-Mb.02-Ma2e4. Note that the time at
which the runs begin to form sinks, $\tsf$, is different in each
case. The {\it top panels} of Fig.\
\ref{fig:SFE_of_t_vs_vinf} then show the SFE, defined as in eq.\
(\ref{eq:SFE_def}). The {\it top left panel} shows \sfeabs, i.e., starting 
from the beginning of the simulation, and up to a total time of 25
Myr. The {\it top right panel} shows the SFE starting from the time at which
sink formation begins in each run, allowing one to read off \sferel. 

 From the {\it top right panel} of Fig.\ \ref{fig:SFE_of_t_vs_vinf} we see
that the mean slope of the curve ~SFE$(t)$, denoted \dSFE, decreases,
although moderately, with increasing $\Msinf$.
It is interesting, however, that all three runs form stars at roughly
the same rate (i.e., the slopes of the curves in the {\it right panel}
of Fig.\ \ref{fig:Mg_Ms_of_t_vs_vinf}, which give the mass accretion
rate onto the sinks, and which we identify with the star formation
rate, SFR, are all similar).\footnote{Note that, because the cloud's
mass is in general not constant in time, eq.\ (\ref{eq:SFE_def}) implies
that $\Mdense \times$\dSFE$ \neq$ SFR.} The decrease in
\dSFE\ is thus due to the larger mass growth rate
of the cloud induced by the larger inflow velocities (Fig.\
\ref{fig:Mg_Ms_of_t_vs_vinf}, {\it top}), not to a smaller SFR. 

\begin{figure}
\includegraphics[width=1.\hsize]{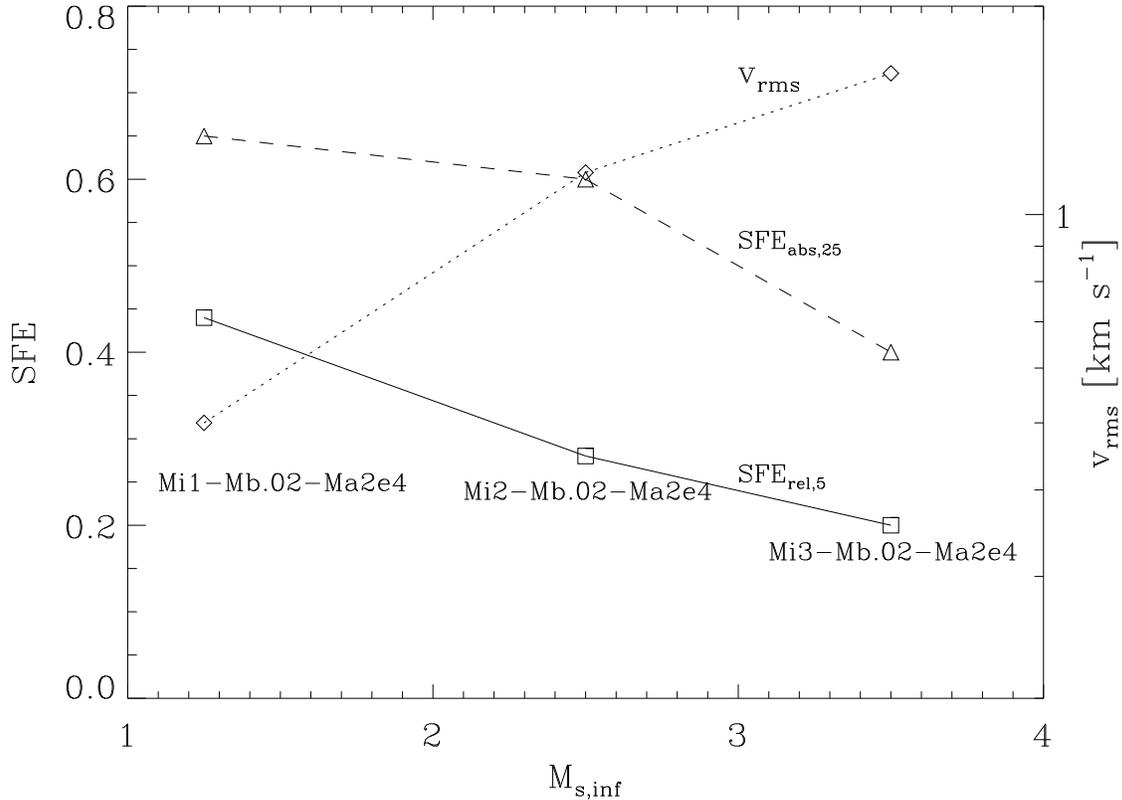}
\caption{Dependence of \sfeabs\ and \sferel\ on the inflow speed for
runs Mi1-Mb.02-Ma2e4, Mi2-Mb.02-Ma2e4, and Mi3-Mb.02-Ma2e4. Both SFEs
are seen to decrease with increasing $\vinf$, apparently due to an
increase of the turbulent velocity dispersion $\vrms$ as the inflow
speed increases, which in turn reduces the SFR.}
\label{fig:sfe_vs_vinf}
\end{figure}

The larger-mass clouds appear to be incapable of 
forming stars at a proportionally larger rate because the larger inflow
velocity produces larger turbulent velocity dispersions $\vrms$ in the
clouds,\footnote{Note that this velocity dispersion, which is the {\it
response} of the flow in its dense regions to the collision of the
inflows, is different from the {\it initial} background turbulent
velocity dispersion applied to the
runs, measured by the parameter $\Mrms$.} which have
been shown to reduce the SFR in simulations of driven turbulence
\citep[e.g.,][]{KHM00, VBK03, MK04, VKB05}. This is illustrated in the
{\it bottom} panel of Fig.\ \ref{fig:SFE_of_t_vs_vinf}, which shows the
evolution of the one-dimensional $\vrms$ perpendicular to the direction of
the inflows for the same three runs.\footnote{Note that the velocity
dispersions in this figure are significantly smaller than those reported in
Figs. 5 (bottom panel) and 9 of Paper I. In that paper, those figures
suffered from a typographical error, having the velocities erroneously
multiplied by one too many factors of the velocity unit, $v_0 = 7.362
\kms$. Thus, the correct velocities in those figures are obtained by
dividing by this factor.} The trends of \sfeabs, \sferel, and $\vrms$
with the inflow speed are summarized in Fig.\
\ref{fig:sfe_vs_vinf}. In this figure, $\vrms$ is measured for Run
Mi1-Mb.02-Ma2e4 at the time when it exhibits a minimum in the {\it
bottom} panel of Fig.\ \ref{fig:SFE_of_t_vs_vinf}, i.e., $t=8$ Myr. This
is equivalent to $\sim 2/3~\tcros$, where $\tcros \equiv \linf/\vinf$ is
the {\it inflow crossing time}, the time taken by the tail of the
inflowing cylinders to reach the collision plane. We do this in an
attempt to capture the true velocity dispersion in the cloud after the
initial transients have ended, but before global collapse sets in. The
latter is indicated by the smooth rise in $\vrms$ for this run during
the time interval $9 \la t \la 23$ Myr. Then, for consistency, we
measure $\vrms$ for the other two runs also at 2/3 of their own
$\tcros$.

\subsection{SFE vs.\ background turbulence strength} \label{sec:SFE_vs_Mrms}

We now consider the response of the SFE to the amplitude of the initial
turbulent velocity field. Note that this field was not originally
intended to produce density condensations on its own, but just to
sufficiently disorganize the inflow velocity field as to trigger the
instabilities that render the cloud turbulent. However, in the cases of
stronger turbulence, we do observe clump formation everywhere in the box
as a result of the initial background turbulence, and not just at the
collision site of the inflows.

\begin{figure}
\includegraphics[width=0.5\hsize]{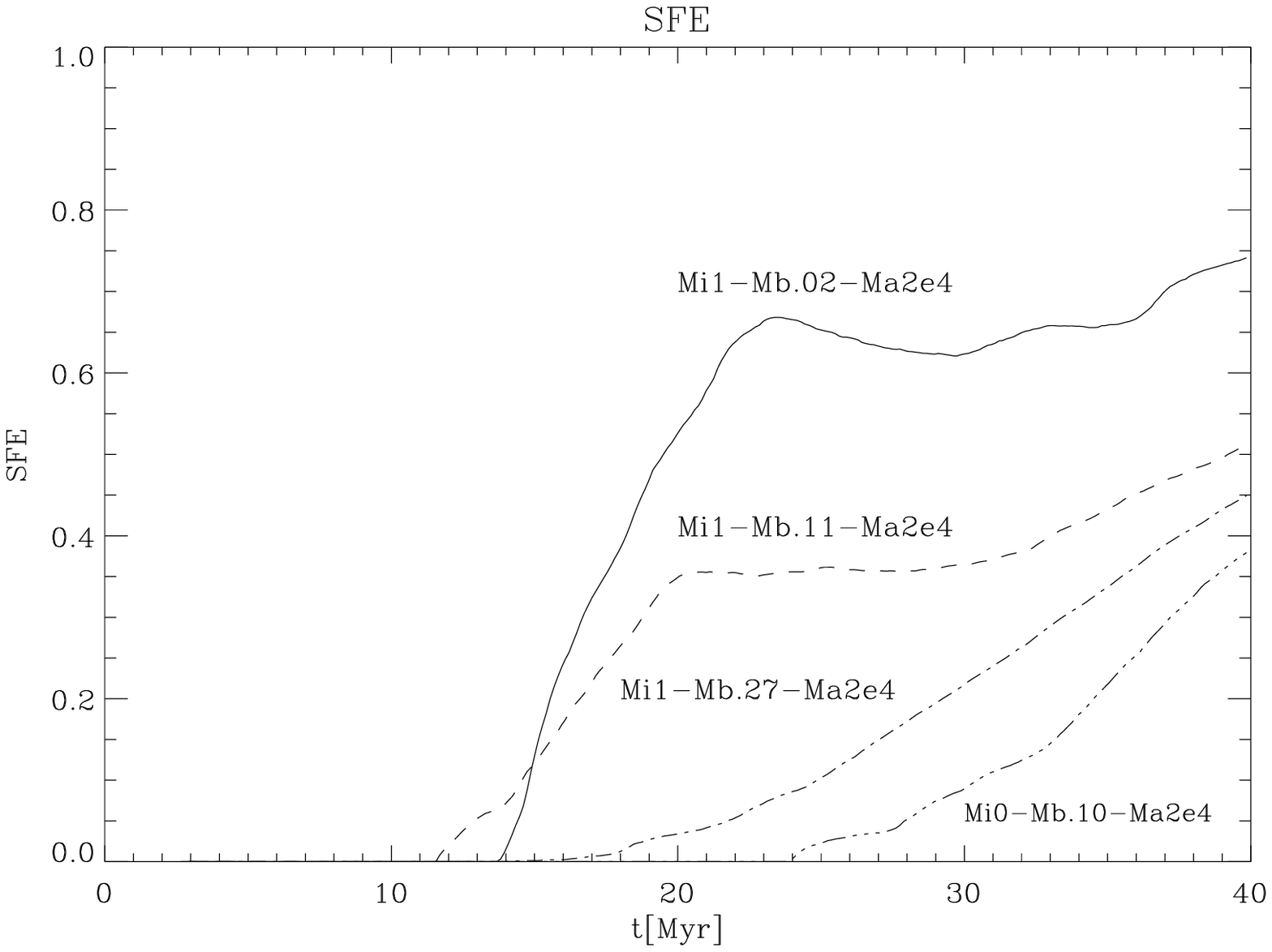}
\includegraphics[width=0.5\hsize]{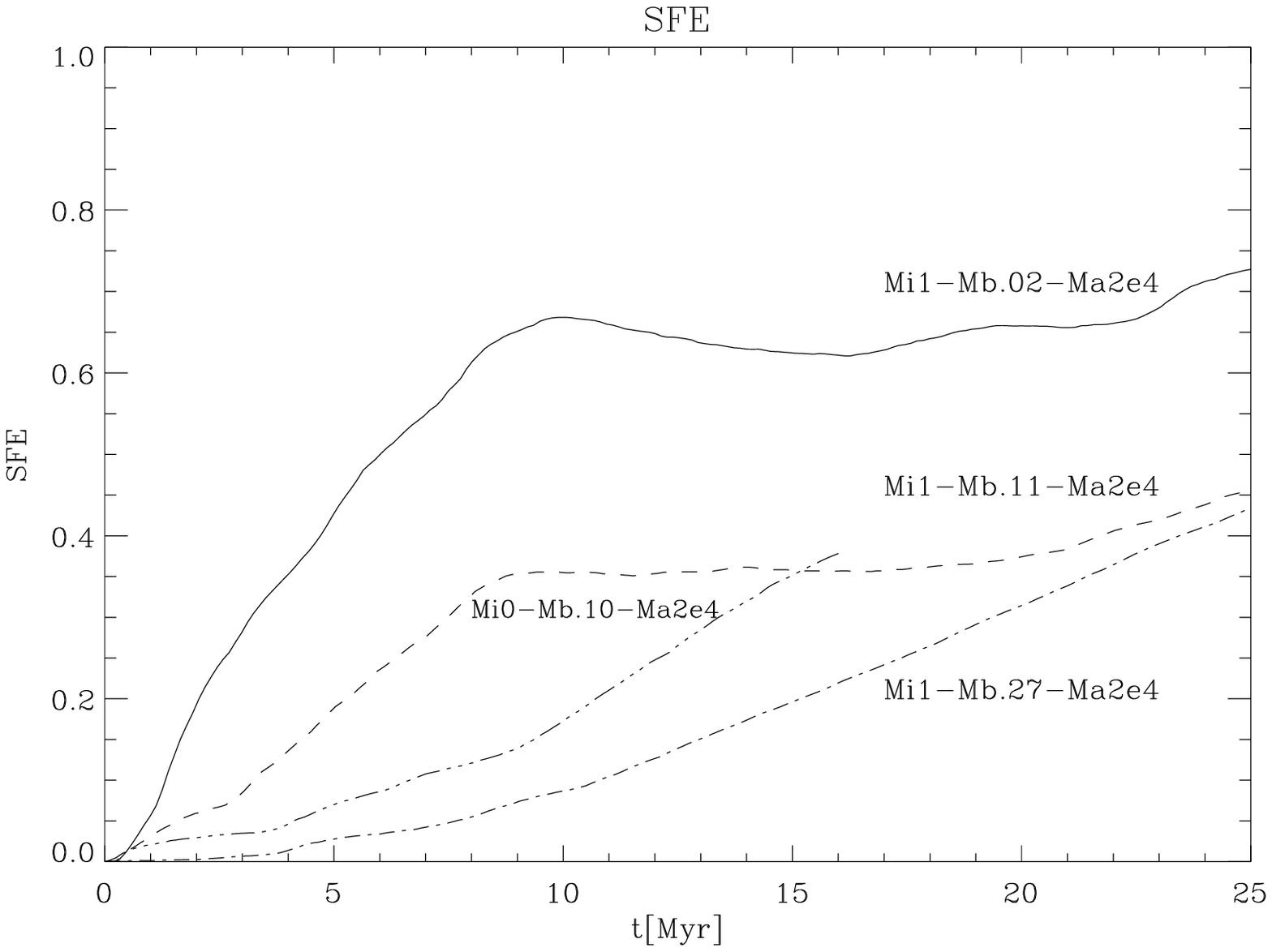}
\caption{Evolution of the ``absolute'' SFE (\sfeabs,
{\it left panel}) and the ``relative'' SFE ({\it right panel}) for four
runs, characterized by various values of the initial background
turbulent Mach number $\Mrms$, indicated by the entry ``Mb.\#\#'' in the
runs' names. Three of the runs have the same inflow Mach number, $\Minf
= 1.25$, while the fourth run has no inflows ($\Minf = 0$), in order to
assess the SFE due exclusively to the initial background turbulence.}
\label{fig:SFEt_vs_vrms}
\end{figure}

Figure \ref{fig:SFEt_vs_vrms} shows the evolution of the SFE for runs
Mi1-Mb.11-Ma2e4, Mi1-Mb.27-Ma2e4, Mi1-Mb.02-Ma2e4 and
Mi0-Mb.10-Ma2e4. The first three runs differ only in the strength of the
initial turbulence, measured by $\Mrms$ (cf.\ Table
\ref{tab:run_params}). The last run has nearly the same value of $\Mrms$
as Mi1-Mb.11-Ma2e4, but with no inflow velocity, in order to assess the
amount of star formation induced solely by the turbulent field, in the
absence of colliding streams.

\begin{figure}
\includegraphics[width=0.45\hsize]{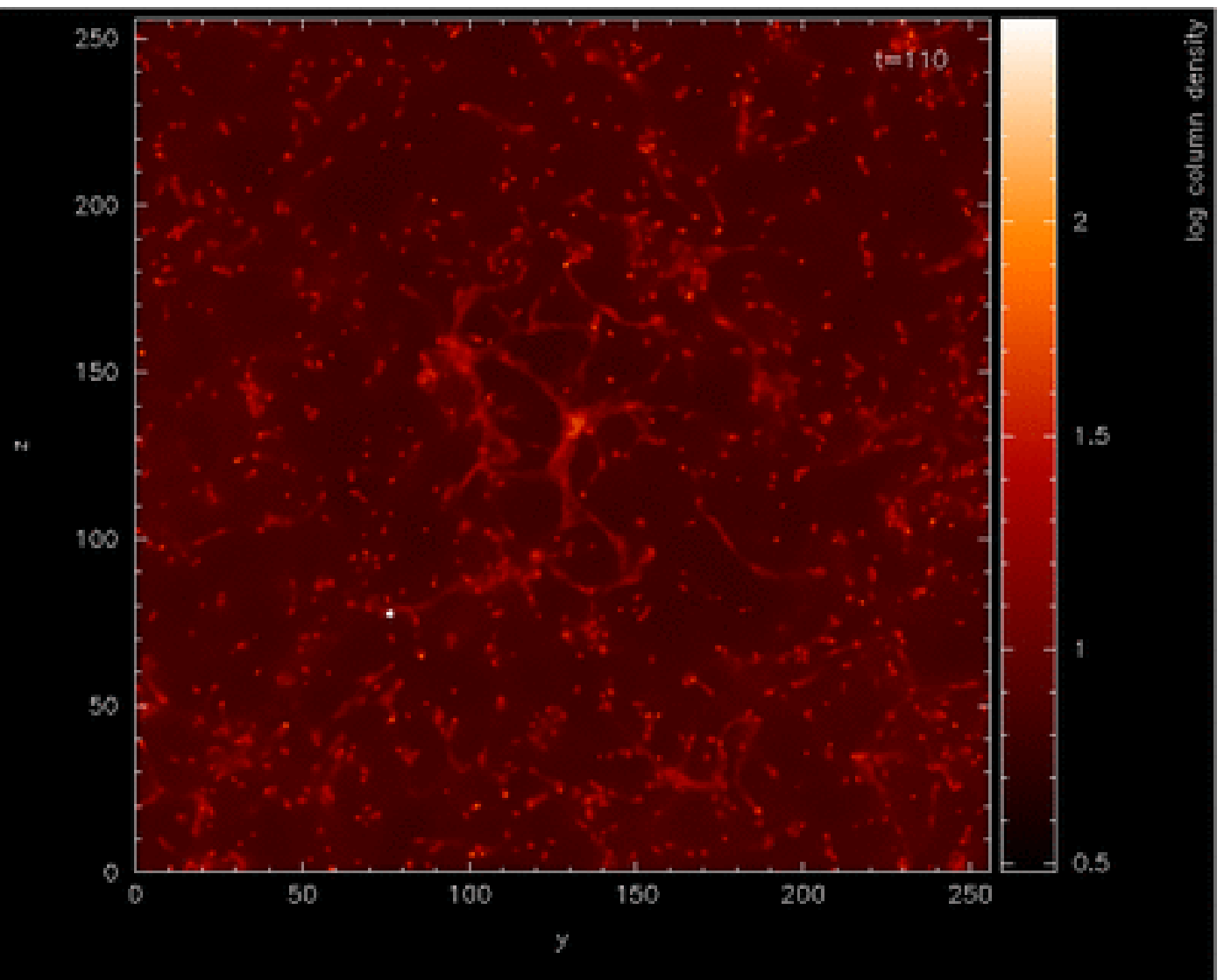}\hspace{0.5cm}
\includegraphics[width=0.45\hsize]{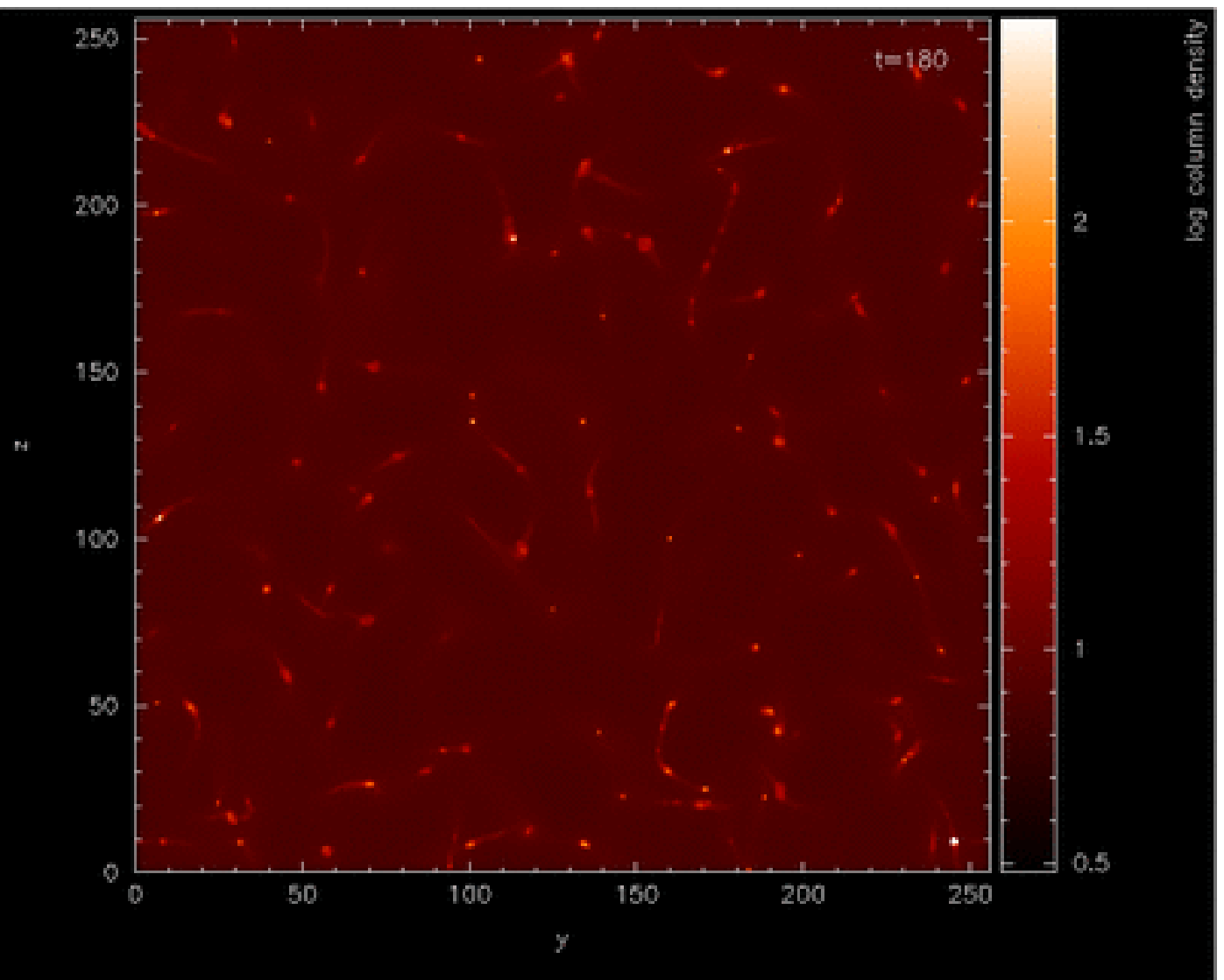}
\caption{Face-on images (in projection) of Mi1-Mb.27-Ma2e4 ({\it left
panel}) and Mi0-Mb.10-Ma2e4 ({\it right panel}) at the time each one is
beginning to form stars. Note the scattered structure, due to the
turbulence forming clumps throughout the numerical box. In run
Mi1-Mb.27-Ma2e4, in which the colliding streams are present, the
turbulence almost completely obliterates the ``main'' cloud formed by
the streams.}
\label{fig:runs3_11_img}
\end{figure}

 From these figures, we see that $\tsf$ becomes longer and \dSFE\
becomes smaller as larger values of $\Mrms$ are 
considered. This appears to be due to the stronger fragmentation
induced by the turbulent velocity field, which in the extreme case of
Mi1-Mb.27-Ma2e4 almost obliterates the inflows, and produces scattered
clumps througout the simulation box, with very little remaining of the
cloud formed by the inflows, as illustrated in Fig.\
\ref{fig:runs3_11_img}, {\it left panel}.

It is also worth noting that run Mi0-Mb.10-Ma2e4, which has no inflows,
has a much larger $\tsf$ and a lower \dSFE\ than run Mi1-Mb.11-Ma2e4,
which differs from the former only in the presence of the inflows. As
shown in Fig.\ \ref{fig:runs3_11_img}, this run also produces scattered
clumps throughout the numerical box, although not as profusely as
Mi1-Mb.27-Ma2e4. So, we conclude that sink formation is still dominated
by the colliding streams in Mi1-Mb.11-Ma2e4, although a small fraction
of the sinks is contributed by the global turbulence.  

\begin{figure}
\includegraphics[width=1.\hsize]{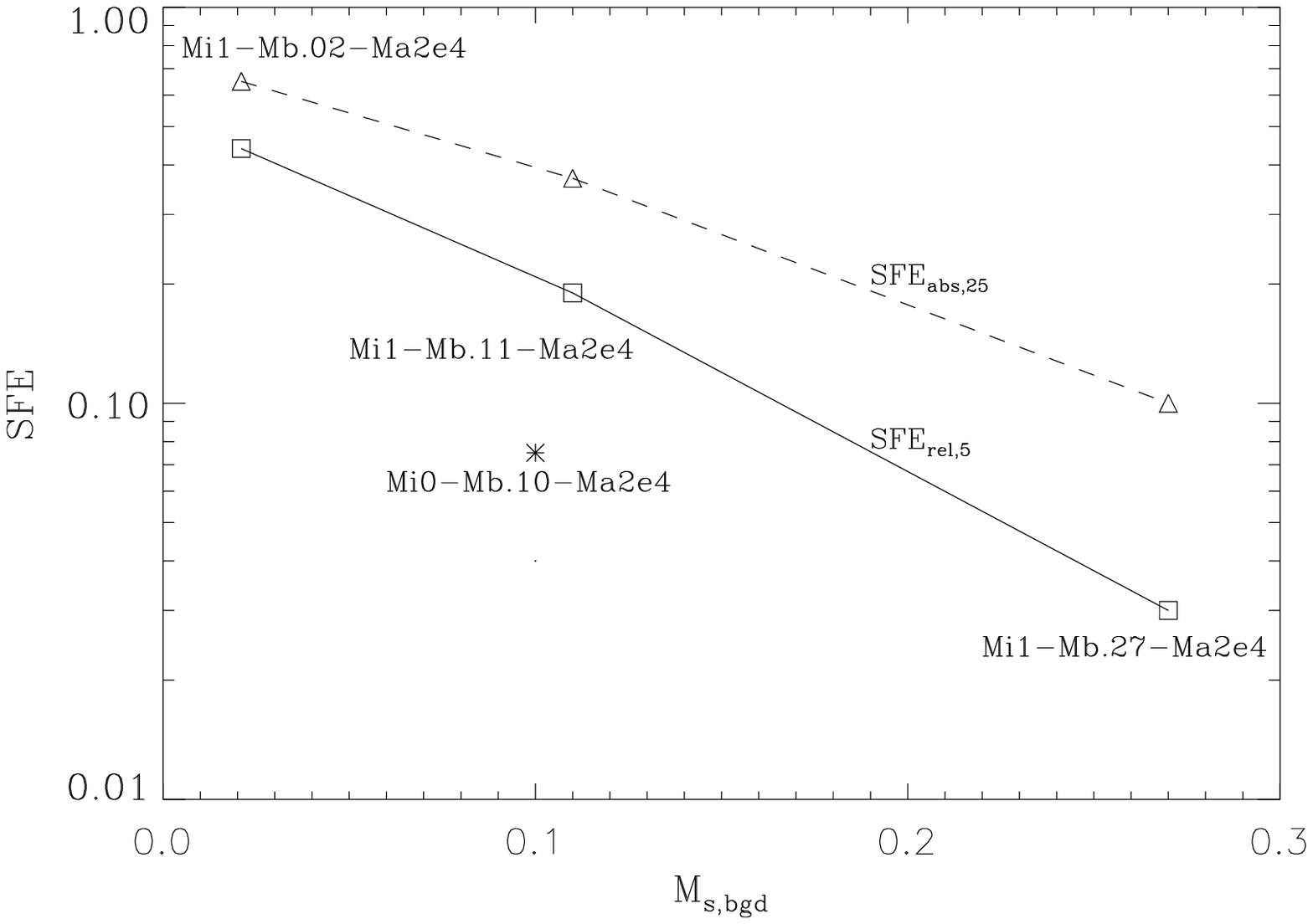}
\caption{Dependence of \sfeabs\ and \sferel\ on the rms Mach number,
$\Mrms$, of the initial (background) turbulent velocity
perturbations. Both indicators are seen to decrease with increasing
$\Mrms$ as a consequence of the progressively stronger fragmentation
induced by the turbulence.}
\label{fig:sfe_vs_Mrms}
\end{figure}

Figure\ \ref{fig:sfe_vs_Mrms} summarises the results of this section. A
clear trend of a decreasing SFE (seen in both \sfeabs\ and \sferel) with
increasing $\Mrms$ is seen, which we interpret as a result of the
reduction of the fragment mass with increasing turbulence strength
\citep[at constant total mass; ][]{BP_etal06}, and of the fact that
smaller-mass fragments tend to have smaller SFEs (\S \ref{sec:SFE_vs_Rinf}).

\subsection{SFE vs.\ inflow mass} \label{sec:SFE_vs_Rinf}

The last dependence of the SFE we analyse is on the mass content
of the colliding inflows. So, we consider inflows of various radii.
However, since a very narrow inflow is necessarily more poorly resolved,
we consider smaller simulation boxes in two of the cases, in order to
better resolve the resulting clouds. Specifically, as shown in Table
\ref{tab:run_params}, $\Rinf$ in run Mi1-Mb.02-Ma5e3 is half that in
Mi1-Mb.02-Ma2e4. Run Mi1-Mb.06-Ma2e3 has $\Rinf$ equal to that of
Mi1-Mb.02-Ma5e3, but half the length, since the numerical box size of
the former is half that of the latter. Finally, Mi1-Mb.06-Ma6e2 has
$\Rinf$ equal to half that of Mi1-Mb.06-Ma2e3. So, the total mass
contained in the inflows of runs Mi1-Mb.02-Ma2e4, Mi1-Mb.02-Ma5e3,
Mi1-Mb.06-Ma2e3 and Mi1-Mb.06-Ma6e2 is, respectively, $2.26 \times
10^4$, $5.64 \times 10^3$, $2.42 \times 10^3$, and $6.04 \times 10^2
\Msun$.

\begin{figure}
\includegraphics[width=0.5\hsize]{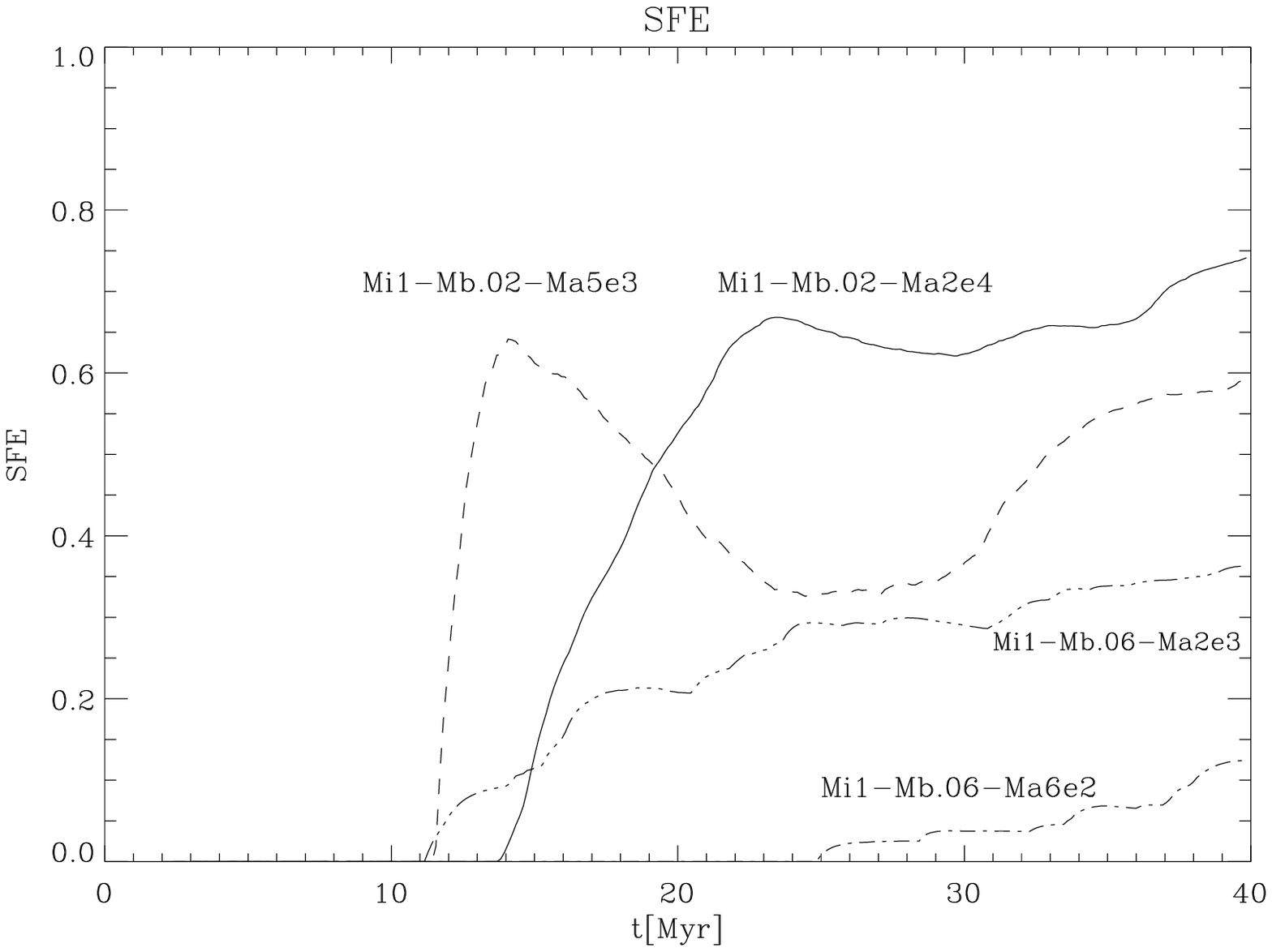}
\includegraphics[width=0.5\hsize]{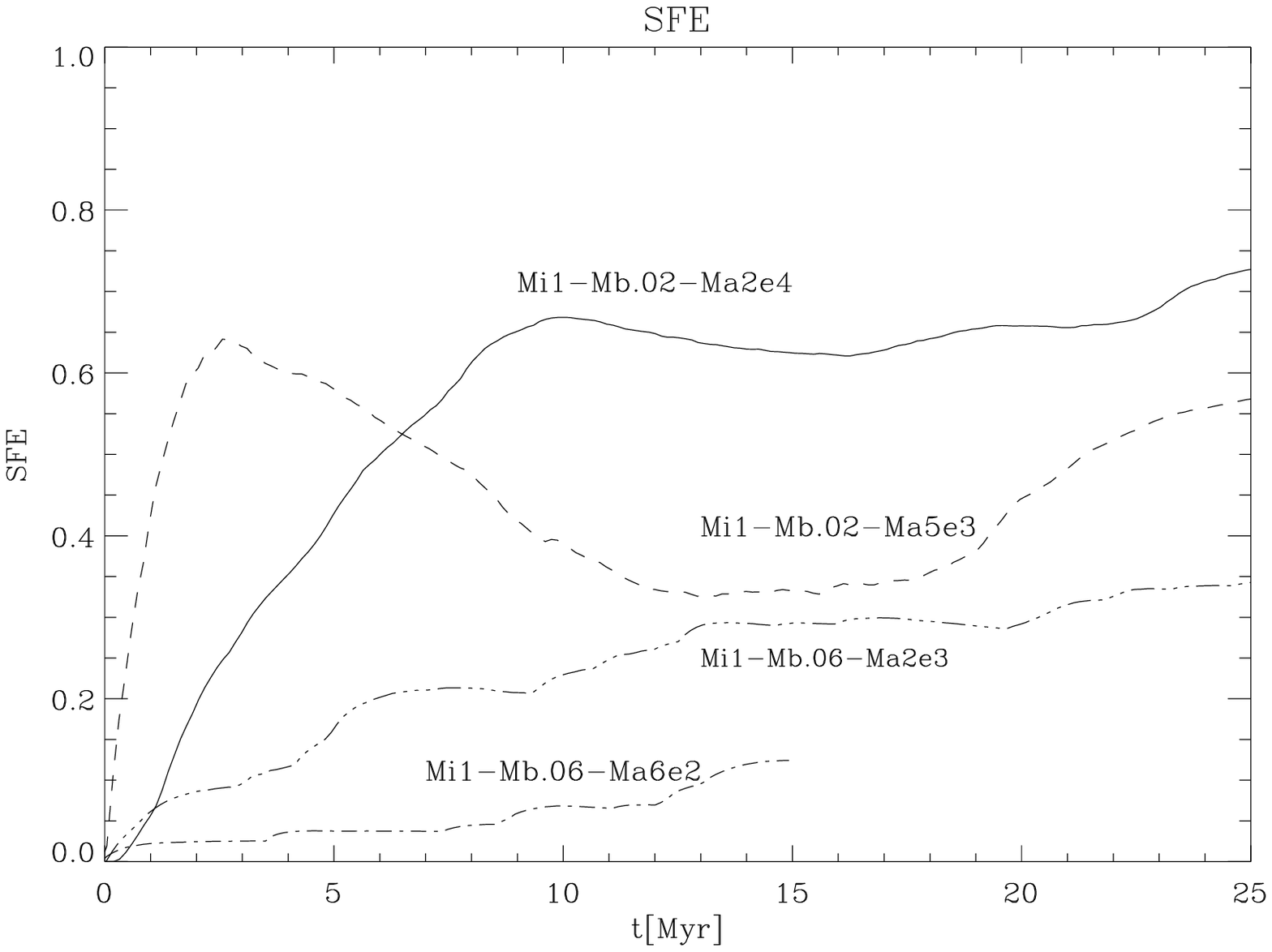}
\caption{Evolution of the ``absolute'' SFE (\sfeabs,
{\it left panel}) and the ``relative'' SFE ({\it right panel}) for runs
Mi1-Mb.02-Ma2e4, Mi1-Mb.02-Ma5e3, Mi1-Mb.06-Ma2e3, and
Mi1-Mb.06-Ma6e2. The radius of the cylindrical inflows for these runs is
respectively 32, 16, 16, and 8 pc. Runs Mi1-Mb.02-Ma2e4 and
Mi1-Mb.02-Ma5e3 are performed in a 256-pc box, with inflow length 112
pc, while runs Mi1-Mb.06-Ma6e2 and Mi1-Mb.06-Ma2e3 are performed in a
128-pc box with the same number of SPH particles (thus being better
resolved) and a 48-pc inflow length.}
\label{fig:SFEt_vs_rcyl}
\end{figure}

\begin{figure}
\includegraphics[width=1.\hsize]{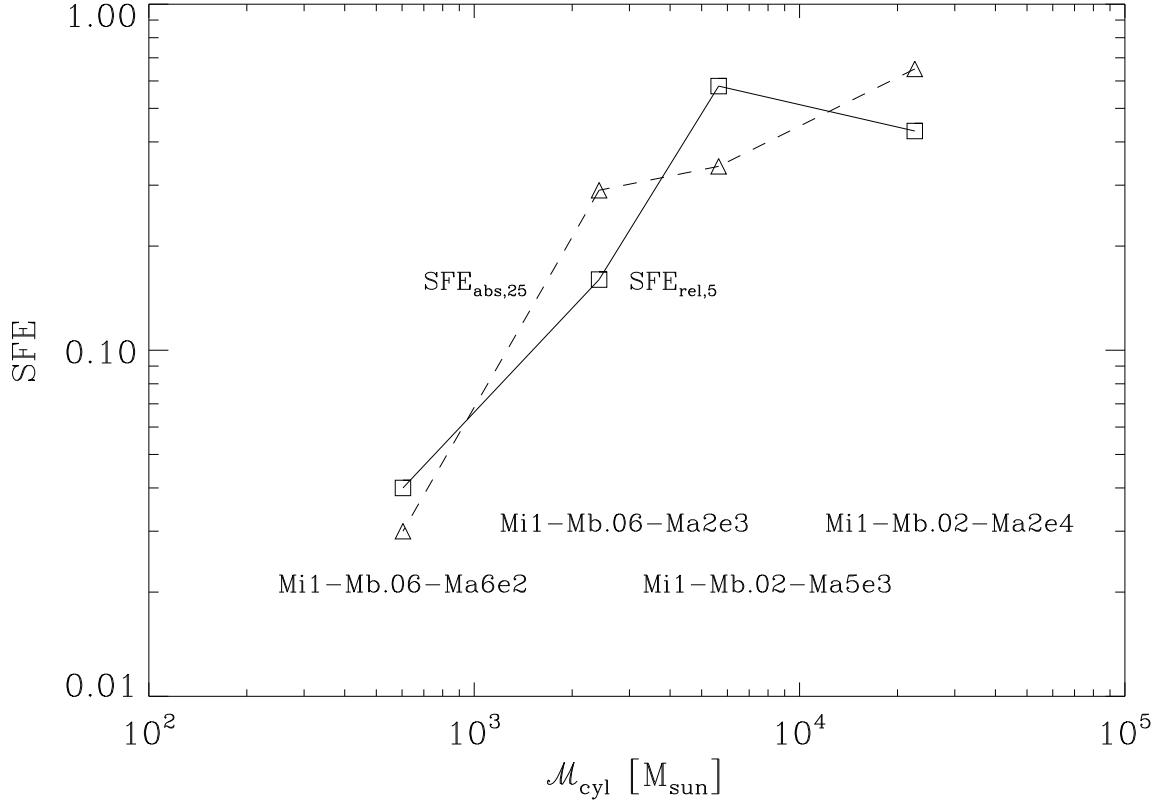}
\caption{Dependence of \sfeabs\ and \sferel\ on the inflows' mass. A
general trend for the SFE to increase with inflow mass is observed.}
\label{fig:sfe_vs_Mcyl}
\end{figure}

Figure \ref{fig:SFEt_vs_rcyl} shows the evolution of \sfeabs\ and
\sferel\ for these runs. We see that there is a general trend for the
SFE, in both its forms, to increase with the total mass involved in the
stream collision. There is only a reversal to this trend in the relative
SFE between runs Mi1-Mb.02-Ma2e4 and Mi1-Mb.02-Ma5e3 because the latter
has a large early maximum of \sferel, although later it decreases, in a
period of mass accumulation in the cloud at low SFR. Other than that,
the trend is general, as shown in Fig.\ \ref{fig:sfe_vs_Mcyl}.

It is important to note that for all runs in this series we used the
same energy injection rate of the turbulence driver. However, runs
Mi1-Mb.06-Ma6e2 and Mi1-Mb.06-Ma2e3, performed in a smaller
computational box, have an initial, background rms turbulent Mach number
that is roughly 2.5 times larger than that of runs Mi1-Mb.02-Ma2e4 and
Mi1-Mb.02-Ma5e3. This may additionally reduce the SFE because of the
additional fragmentation it produces, but we see that
the trend of the SFE to decrease with decreasing inflow mass holds
generally even at the same physical box size, so the result appears
robust.

The trend discused above can be put in the context of the theory of
\citet[][hereafter KM05]{KM05} for the \sfrff. This theory predicts a
dependence of the \sfrff\ on the virial parameter $\alpha$ and the rms
Mach number $\Ms$. Here,
\begin{equation}
\alpha \equiv 2 \Ek/|\Eg|,
\label{eq:alpha_def}
\end{equation}
where $\Ek = {\cal M} \Delta v^2/2$ is the cloud's kinetic energy,
${\cal M}$ is the cloud's mass, $\Delta v$ is its rms turbulent
velocity dispersion, and $\Eg$ is the cloud's gravitational energy.

For our flattened clouds, we compute the gravitational energy assuming
they can be approximated as infinitely thin, uniform disks of radius
$R$, and write
\begin{equation}
\Eg = \int_A \Sigma \phi~d^2 x = 2 \pi \Sigma \int_0^R r \phi(r) dr,
\label{eq:Eg_def}
\end{equation}
where $A$ is the area of the disk, $\Sigma$ is
the (uniform) surface density, and $\phi$ is the gravitational potential.
In our case, the latter is given by \citep{WM42, BH04}
\begin{equation}
\phi(r) = -4 G \Sigma R E(r/R),
\label{eq:pot}
\end{equation}
where $E$ is the second complete elliptic integral. Thus, the
gravitational energy is
\begin{eqnarray}
\Eg &=& - 8 \pi G \Sigma^2 R \int_0^R r E(r/R) dr = -8 \pi G \Sigma^2 R^3
\int_0^1 x E(x) dx \nonumber \\
    &=& - 8 \pi \left(\frac{28}{45}\right) G \Sigma^2 R^3.
\label{eq:Eg_calc}
\end{eqnarray}

To compute the kinetic energy of the clouds, we note that the relevant
velocity dispersion is the one produced in the clouds as a consequence of
the inflow collision \citep{Heitsch_etal05, VS_etal06}, rather than the
initial background turbulent velocity, which is much smaller. Since all
four simulations analysed in this section have the same $\vinf$, we use
the value of $\vrms$ measured for Run Mi1-Mb.02-Ma2e4 in \S\
\ref{sec:SFE_vs_vinf} (Fig. \ref{fig:sfe_vs_vinf}) for all of them,
namely $\vrms =0.5 \kms$. Noting that this value is a one-dimensional
velocity dispersion, we take $\Delta v = \sqrt{3}~\vrms$.

\begin{figure}
\includegraphics[width=1.\hsize]{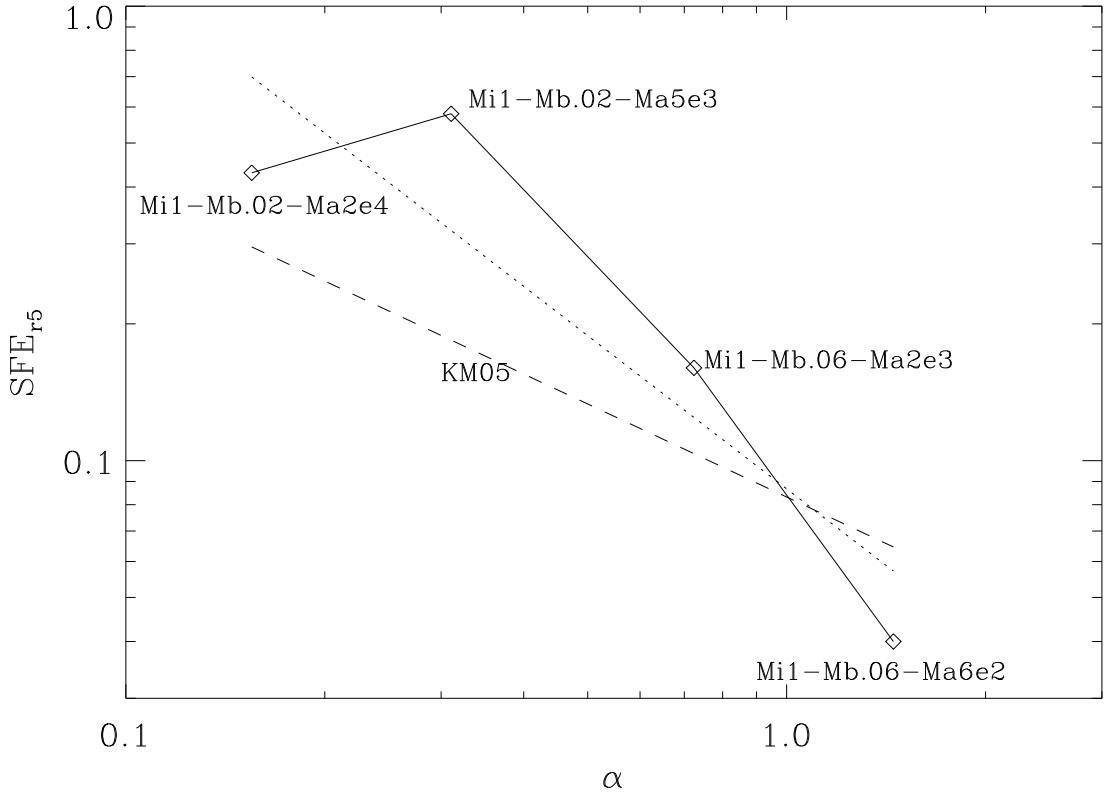}
\caption{Dependence of \sferel\ on the virial parameter $\alpha$. The
straight {\it dotted} line shows a least squares fit to the simulation
data, with slope $-1.12$, while the {\it dashed} line shows the result
from \citet{KM05}, with slope $-0.68$.}
\label{fig:sfe_vs_alpha}
\end{figure}

We finally obtain, from equations (\ref{eq:alpha_def}) and
(\ref{eq:Eg_calc}), 
\begin{equation}
\alpha = \frac{224 \pi R \Delta v^2}{45 G M}.
\label{eq:alpha_calc}
\end{equation}
Figure \ref{fig:sfe_vs_alpha} shows the results of this exercise. The
{\it solid} line shows the simulation data, while the straight {\it dotted}
line shows a least squares fit to them. The {\it dashed} line shows the
result from \citet{KM05}, given by
\begin{equation}
\hbox{\sfrff} \approx 0.014 \left(\frac{\alpha}{1.3} \right)^{-0.68} \left(
\frac{\Ms}{100} \right)^{-0.32},
\label{eq:SFE_KM05}
\end{equation}
where we have taken $\Ms = \Delta v/\cs$.
We see that the prediction by KM05, although being numerically within
the same range as the data, exhibits a significantly shallower slope.
Specifically, the fit to our data has a slope $-1.12 \pm 0.37$, where
the uncertainty is the $1\sigma$ error of the fit, while the slope of
the KM05 prediction, $-0.68$, lies beyond this error. We discuss this
result further in \S \ref{sec:conclusions}.

\section{Summary and discussion} \label{sec:conclusions}

In this paper, we have considered the scenario of molecular cloud
formation by WNM stream collisions, and investigated the dependence of
the SFE on three parameters of this scenario, namely the inflow speed,
the rms Mach number of the background medium, and the total mass
contained in the inflows. Since the SFE, defined as in eq.\
(\ref{eq:SFE_def}), is a time-dependent function because the cloud
continues to accrete mass from the WNM while it forms stars, we have
considered two estimators of its time integral, namely the absolute SFE after
25 Myr from the start of the simulation, \sfeabs, and the ``relative''
SFE, 5 Myr after the onset of SF in the cloud, denoted \sferel.

We have found a wide range of values of these estimators as we vary the
parameters of the simulations. In general, the SFE decreases, although
moderately, with increasing inflow velocity. In particular, \sferel\
decreases from $\sim 0.4$ to $\sim 0.2$ as $\Msinf$ increases from 1.25
to 3.5. Note that the runs in this case have similar SFRs, and the
decrease in the SFE and \dSFE\ is due to the faster increase in cloud
mass rather than to a decrease in the SFR. That the SFR is similar in all
three runs, in spite of the larger gas mass is probably due to the
larger turbulent velocity dispersion in the dense gas caused by the
larger inflow speed, which tends to inhibit the SFR, thus compensating
the tendency to have a larger SFR due to the larger cloud masses.

Similarly, the SFE decreases with increasing background turbulence
strength, as the latter progressively takes a dominant role in the
production of the dense gas but, due to the relatively small scales at
which the turbulence is excited, the clouds and clumps formed by it are
significantly smaller than the cloud formed by the coherent stream
collision. In this case, \sferel\ decreases from $\sim 0.4$ to $\sim
0.03$ as the rms Mach number of the background turbulence increases from
$\sim 0.02$ to $\sim 0.3$.

Finally, the SFE in general decreases with decreasing mass of the
inflows (at constant $\vinf$). This may be a consequence of the fact
that clouds formed by the collision of our inflows always have roughly
the same density, temperature, and velocity dispersion, so smaller
clouds are more weakly gravitationally bound, a condition known to
decrease the SFE \citep{Clark_etal05}. The end result is that
\sferel\ decreases from $\sim 0.4$ to $\sim 0.03$ as the mass in the
colliding streams decreases from $\sim 2.3 \times 10^4 \Msun$ to $\sim
600 \Msun$. It is important to stress that this result is not at odds
with the well known fact that the SFE {\it increases} as the object mass
decreases from the mass scale of a giant molecular cloud (${\cal M} \sim
10^4$ -- $10^6 \Msun$, SFE $\sim 0.02$; Myers et al.\ 1986) to that of a
cluster-forming core (${\cal M} \sim 10^3 \Msun$; SFE $\sim 0.3$ -- 0.5,
Lada \& Lada 2003), because in this case the cores' mean densities are
much larger than those of the GMCs, while in our case the mean densities
of the various clouds are always comparable. Thus, our clouds do not
conform to Larson's (1981) density-size scaling.

The latter results, expressed in terms of the virial parameter $\alpha$,
exhibit partial agreement with the prediction by KM05 for the dependence
of the SFE after a free-fall time (what those authors called the star
formation rate per free-fall time, or \sfrff), which is directly
comparable to our \sferel. Although their prediction, without any
rescaling, lies in the same range of values as our observed
efficiencies, it contains a much shallower dependence on $\alpha$ than we
observe. This may be due to the fact that those authors assumed that the
clouds were supported by turbulent pressure, possibly provided by
stellar feedback, while our simulations lack such support. However, the
notion of turbulent support has been questioned recently by
\citet{VS_etal08}, and the low observed efficiency may be the result of
the {\it dispersal} (rather than support) of the parent cloud by its
stellar products \citep{HBB01}. More work is needed to determine the
causes of the discrepancy. However, it is noteworthy that Run
Mi1-Mb.06-Ma2e3, which has a value of $\alpha$ closest to unity, has a
reasonably realistic value of \sferel$ \sim 4$\%.

We conclude that the SFE, even in the absence of further agents such as
magnetic fields and stellar feedback, is a highly sensitive function of
the parameters of the cloud formation process, and may be responsible
for significant intrinsic scatter in observational determinations of the
SFE.

\section*{Acknowledgments}

The numerical simulations were performed in the cluster at CRyA-UNAM
acquired with CONACYT grants 36571-E and 47366-F to E.V.-S. G.C.G.\
acknowledges financial support from grants IN106809 (UNAM-PAPIIT) and
J50402-F (CONACYT). A.-K.J.\ acknowledges
support by the Human Resources and Mobility Programme of the European
Community under the contract MEIF-CT-2006-039569.
%
%

\newpage

\begin{table}
\caption{Run parameters}
\label{tab:run_params}
\begin{center}
\begin{tabular}{|c|c|c|c|c|c|c|c|c|c||c|}
\hline
Run & Run& $\Lbox$ & $\linf$ & $\vinf$ & $\Msinf$ & $\Mrms$    & $\Rinf$ &
$\Mbox$ & $\Minf$ & ${\cal M}_{\rm part}$\\
number & name & [pc]    &  [pc]  &$[\kms]$&         &           & [pc]  & $[\Msun]$ & $[\Msun]$ & $[\Msun]$ \\
\hline
1 & Mi1-Mb.11-Ma2e4 & 256      & 112       &  9.20       &    1.25 &  0.11        & 32   & 
$5.25\times10^5$  &  $2.26 \times 10^4$  &  $0.32$
\\
2 & Mi2-Mb.11-Ma2e4 & 256      & 112       &  18.41     &     2.50 &  0.11       &  32    &
$5.25\times10^5$   &  $2.26 \times 10^4$  &  $0.32$
\\
3 & Mi1-Mb.27-Ma2e4 & 256      & 112       & 9.20       &     1.25 &  0.27       &  32    &
$5.25\times10^5$  &  $2.26 \times 10^4$  &  $0.32$
\\
4 & Mi1-Mb.02-Ma2e4 & 256      &112        & 9.20       &    1.25 &   0.021      &  32   &
$5.25\times10^5$  &  $2.26 \times 10^4$  &  $0.32$
\\
5 & Mi2-Mb.02-Ma2e4 & 256      &112        & 18.41      &   2.50  &   0.024      &  32  &
$5.25\times10^5$  &  $2.26 \times 10^4$  &  $0.32$
\\
6 & Mi3-Mb.02-Ma2e4 &256       &112        & 25.77      &  3.50   &   0.025      &  32  &
$5.25\times10^5$  &  $2.26 \times 10^4$  &  $0.32$
\\
8 & Mi1-Mb.02-Ma5e3 &256       &112       & 9.20        &  1.25  &  0.020        &  16 &
$5.25\times10^5$  &  $5.64 \times 10^3$  &  $0.32$
\\
10 & Mi1-Mb.06-Ma6e2 &128     &48         & 9.20        &1.25    & 0.057         &  8      &
$6.57\times10^4$  &  $6.04 \times 10^2$  &  $0.04$
\\
11 & Mi0-Mb.10-Ma2e4 &256     &112         & 0.          &0.     & 0.10          &  32     &
$5.25\times10^5$  &  $2.26 \times 10^4$  &  $0.32$
\\
12 & Mi1-Mb.06-Ma2e3 &128     &48         & 9.20        &1.25    & 0.058         &  16     &
$6.57\times10^4$  &  $2.42 \times 10^3$  &  $0.04$
\\
\hline
 
\end{tabular}
\end{center}
\end{table}


\label{lastpage} 

\begin{thebibliography}{99}


\bibitem[Ballesteros-Paredes et al.(2006)]{BP_etal06} 
Ballesteros-Paredes, J., Gazol, A., Kim, J., Klessen, R.~S., Jappsen, 
A.-K., \& Tejero, E.\ 2006, ApJ, 637, 384 

\bibitem[Ballesteros-Paredes et al.(2007)]{BP_etal07} 
Ballesteros-Paredes, J., Klessen, R.~S., Mac Low, M.-M., 
\& Vazquez-Semadeni, E.\ 2007, Protostars and Planets V, 63 

\bibitem[Bate \& Burkert(1997)]{BB97} Bate, M.~R., \& Burkert, A.\ 1997,
MNRAS, 288, 1060 

\bibitem[Burkert \& Hartmann(2004)]{BH04} Burkert, A., \& Hartmann, L.\
2004, ApJ, 616, 288 

\bibitem[Clark et al.(2005)]{Clark_etal05} Clark, P.~C., Bonnell, 
I.~A., Zinnecker, H., \& Bate, M.~R.\ 2005, MNRAS, 359, 809 

\bibitem[Field(1965)]{Field65} Field, G. B., 1965, ApJ, 142, 531

\bibitem[Franco, Shore \& Tenorio-Tagle(1994)]{FST94} Franco, J.
Shore, S. N., \& Tenorio-Tagle, G. 1994, ApJ, 436, 795

\bibitem[Hartmann, \BP\ \& Bergin(2001)]{HBB01}
Hartmann, L., Ballesteros-Paredes, J., \& Bergin, E. A. 2001, ApJ, 562, 852

\bibitem[\protect\citeauthoryear{Heitsch et al.}{2005}]{Heitsch_etal05}
Heitsch, F., Burkert, A., Hartmann, L., Slyz, A. D. \& Devriendt,
J. E. G. 2005, ApJ, 633, L113

\bibitem[\protect\citeauthoryear{Heitsch et al.}{2006}]{Heitsch_etal06}
Heitsch, F., Slyz, A., Devriendt, J.,  Hartmann, L., \& Burkert, A.
2006, ApJ, 648, 1052

\bibitem[\protect\citeauthoryear{{Heitsch}, {Hartmann}, {Slyz}, {Devriendt} \&
  {Burkert}}{{Heitsch} et~al.}{2008}]{Heitsch_etal08}
{Heitsch} F.,  {Hartmann} L.~W.,  {Slyz} A.~D.,  {Devriendt} J.~E.~G.,
  {Burkert} A.,  2008, ApJ, 674, 316

\bibitem[\protect\citeauthoryear{Hennebelle \& P\'erault}{Hennebelle \&
  P\'erault}{1999}]{HP99} Hennebelle, P., P\'erault, M., 1999, A\&A,
  351, 309

\bibitem[\protect\citeauthoryear{{Hennebelle}, {Banerjee},
{V\'azquez-Semadeni}, {Klessen} \& {Audit}}{{Hennebelle}
et~al.}{2008}]{HBVKA08} {Hennebelle} P., {Banerjee} R.,
{V\'azquez-Semadeni} E., {Klessen} R., {Audit} E., 2008, A\&A, 486, L43

\bibitem[\protect\citeauthoryear{Hennebelle, Mac Low, \&
\VS}{2008}]{HMV08} Hennebelle, P., Mac Low, M.-M., \& \VS\ 2007, in
Structure Formation in the Universe: Galaxies, Stars, Planets,
ed. G. Chabrier (Cambridge: Cambridge University Press), in press
(arXiv:0711.2417) 

\bibitem[Jappsen et al.(2005)]{Jappsen_etal05} Jappsen, A.-K.,
Klessen, R.~S., Larson, R.~B., Li, Y., and Mac Low, M.-M. 2005, A\&A, 435, 611

\bibitem[Klessen, Heitsch \& Mac Low(2000)]{KHM00} Klessen, R. S.,
Heitsch, F., \& MacLow, M. M. 2000, ApJ, 535, 887

\bibitem[Koyama \& Inutsuka(2000)]{KI00} Koyama, H. \& Inutsuka,
S.-I. 2000, ApJ, 532, 980

\bibitem[Koyama \& Inutsuka(2002)]{KI02}
Koyama, H. \& Inutsuka, S.-I. 2002, ApJ, 564, L97

\bibitem[Kroupa(2001)]{Kroupa01} Kroupa, P. 2001, MNRAS, 322, 231

\bibitem[Krumholz \& McKee(2005)]{KM05} Krumholz, M.~R., \& McKee,
C.~F.\ 2005, ApJ, 630, 250 (KM05) 

\bibitem[Heitsch \& Hartmann(2008)]{HH08} Heitsch, F., \&
Hartmann, L.\ 2008, ApJ, 689, 290

\bibitem[Lada \& Lada(2003)]{LL03} Lada, C.~J., \& Lada,
E.~A.\ 2003, ARAA, 41, 57  

\bibitem[Larson(1981)]{Larson81} Larson, R.~B.\ 1981, MNRAS, 
194, 809 

\bibitem[Mac Low \& Klessen(2004)]{MK04} Mac Low, M.-M., \& Klessen,
R. S. 2004, Rev. Mod. Phys., 76, 125

\bibitem[McKee \& Ostriker(2007)]{MO07} McKee, C.~F., \&
Ostriker, E.~C.\ 2007, ARAA, 45, 565  

\bibitem[Myers et al.(1986)]{Myers_etal86} Myers, P. C., Dame, T. M.,
Thaddeus, P., Cohen, R. S., Silverberg, R. F., Dwek, E. \& Hauser,
M. G. 1986, ApJ, 301, 398

\bibitem[Nakamura \& Li(2005)]{NL05} Nakamura, F., \& Li,
Z.-Y.\ 2005, ApJ, 631, 411  

\bibitem[Springel(2005)]{Springel05} Springel, V.\ 2005, MNRAS, 
364, 1105

\bibitem[V\'azquez-Semadeni(2007)]{VS07}
V\'azquez-Semadeni, E. 2007, in Triggered Star Formation in a Turbulent
ISM, eds. B. G. Elmegreen \& J. Palous (Cambridge: Cambridge Univ. Press),
292

\bibitem[\VS, \BP\ \& Klessen(2003)]{VBK03} V\'azquez-Semadeni, E.,
Ballesteros-Paredes, J. \& Klessen, R. 2003, ApJ, 585, L131

\bibitem[\VS, Kim \& \BP(2005)]{VKB05} \VS, E., Kim, J. \& \BP,
J. 2005, ApJ, 630, L49

\bibitem[\protect\citeauthoryear{\VS\ et al.}{2006}]{VS_etal06}
V\'azquez-Semadeni, E., Ryu, D., Passot, T., Gonz\'alez, R. F., \&
Gazol, A.., 2006, ApJ, 643, 245 

\bibitem[\protect\citeauthoryear{\VS\ et al.}{2007}]{VS_etal07} \VS, E.,
G\'omez, G. C., Jappsen, A. K., 
Ballesteros-Paredes, J., Gonz\'alez, R. F., \& Klessen, R. S. 2007,
ApJ, 657, 870 (Paper I)

\bibitem[V\'azquez-Semadeni et al.(2008)]{VS_etal08} 
V{\'a}zquez-Semadeni, E., Gonz{\'a}lez, R.~F., Ballesteros-Paredes, J., 
Gazol, A., \& Kim, J.\ 2008, MNRAS, 390, 769 

\bibitem[Vishniac(1994)]{Vishniac94} Vishniac, E. T. 1994, ApJ, 428, 186

\bibitem[\protect\citeauthoryear{Walder \& Folini}{2000}]{WF00} Walder,
R. \& Folini, D. 2000, ApSS, 274, 343

\bibitem[Wyse \& Mayall(1942)]{WM42} Wyse, A.~B., \& Mayall, N.~U.\
1942, ApJ, 95, 24 

\end{thebibliography}
\end{document}